\documentclass[11pt,a4paper,english,superscriptaddress,aps]{revtex4}
\usepackage[dvips]{graphicx}
\usepackage[all]{xy}
\usepackage[latin1]{inputenc}

\newcommand{\be}{\begin{equation}}
\newcommand{\ee}{\end{equation}}
\newcommand{\ben}{\begin{eqnarray}}
\newcommand{\een}{\end{eqnarray}}

\begin{document}

\title{The   Noncommutative Anandan's Quantum  Phase}
\author{E. Passos}
\affiliation{{ Departamento de
F\'{\i}sica, Universidade Federal da Para\'\i ba, Caixa Postal 5008, 58051-970,
Jo\~ao Pessoa, PB, Brazil}}
\email{lrr,passos,jroberto,furtado@fisica.ufpb.br}
\author{L. R. Ribeiro}
\affiliation{{ Departamento de
F\'{\i}sica, Universidade Federal da Para\'\i ba, Caixa Postal 5008, 58051-970,
Jo\~ao Pessoa, PB, Brazil}}
\email{lrr,passos,jroberto,furtado@fisica.ufpb.br}
\author{C. Furtado}
\affiliation{{ Departamento de
F\'{\i}sica, Universidade Federal da Para\'\i ba, Caixa Postal 5008, 58051-970,
Jo\~ao Pessoa, PB, Brazil}}
\email{lrr,passos,jroberto,furtado@fisica.ufpb.br}
\author{J. R. Nascimento}
\affiliation{{ Departamento de
F\'{\i}sica, Universidade Federal da Para\'\i ba, Caixa Postal 5008, 58051-970,
Jo\~ao Pessoa, PB, Brazil}}
\affiliation{Instituto de F\'\i sica, Universidade de S\~ao Paulo\\
Caixa Postal 66318, 05315-970, S\~ao Paulo, SP, Brazil}
\email{jroberto@fma.if.usp.br}

\begin{abstract}
In this work we study the noncommutative nonrelativistic quantum
dynamics of a neutral particle, that possesses permanent magnetic
and electric dipole moments, in the presence of an external electric
and magnetic fields. We use the Foldy-Wouthuysen transformation of
the Dirac spinor with a non-minimal coupling to obtain the
nonrelativistic limit. In this limit, we study the noncommutative
quantum dynamics and obtain the noncommutative Anandan's geometric
phase. We analyze the situation  where magnetic dipole moment of the
particle is zero and we obtain the noncommutative version of the
He-McKellar-Wilkens effect. We demonstrate that this phase in the
noncommutative case is a geometric dispersive phase. We also
investigate this geometric phase considering the noncommutativity in
the phase space and the Anandan's phase is obtained.
\end{abstract}

\maketitle
\section{Introduction}
In 1959 Aharonov and Bohm \cite{aha} demonstrated that a quantum
charge circulating a magnetic flux tube acquires a quantum
topological phase. This effect was observed experimentally by
Chambers \cite{prl:cham, pes1}. Aharonov and Casher showed that a
particle with a magnetic moment moving in an electric field
accumulates a quantum phase \cite{cas}, which has been observed in a
neutron interferometer \cite{cim} and in a neutral atomic Ramsey
interferometer \cite{san}.

He and McKellar \cite{mac}, and Wilkens \cite{wil} independently,
have predicted the existence of a quantum phase that an electric
dipole acquires while is circulating around and parallel to a line
of magnetic monopoles. A  simple practical experimental
configuration to test this phase was proposed by Wei, Han and Wei
\cite{wei}, where the electric field of a charged wire polarizes a
neutral atom and an uniform magnetic field is applied parallel to
the wire.

In a recent article, a topological phase effect was proposed by
Anandan \cite{prlan} which describes a unified and fully
relativistic covariant treatment of the interaction between a
particle with permanent electric and magnetic dipole moments and an
electromagnetic field. This problem has been  investigated in the
non-relativistic quantum mechanics  by Anandan \cite{plaan} and one
of the authors \cite{furtpra}.

Recently, the interest in the study of physics in noncommutative
space has attracted much interest in several areas of physics
\cite{nekra}. Noncommutative field theories are related to  M-theory
\cite{con}, string theory \cite{sei} and quantum Hall effect
\cite{sus,jela,basu}. In quantum mechanics a great number of
problems have been investigated in the case of the noncommutative
space-time. Some important results obtained are related with
geometric phases such as: Aharonov-Bohm effect \cite{1,2,3,4,5},
Aharonov--Casher effect \cite{mirza,kangepjc} and Berry's quantum
phase \cite{alavi,ghos} and others involving dynamics of dipoles
\cite{cal}. In this paper we analyze the noncommutative quantum
topological phase effect proposed by Anandan for a quantum particle
with permanent magnetic and electric dipole moments in the presence
of an external electric and magnetic fields, and we study the
appearance of a geometrical quantum phase in their dynamics. We
investigate the non-relativistic geometric phase, proposed by
Anandan, for a quantum particle with permanent magnetic and electric
dipole moments in the presence of external electric and magnetic
fields in the noncommutative quantum mechanics. We also investigate
the He-Mckellar-Wilkens phase in noncommutative space.

This article is organized in the following way: in the next section,
we discuss the quantum dynamics of a neutral particle in the
presence of external electromagnetic field. In the section \ref{2},
the noncommutative  non-relativistic quantum  dynamics of quantum
dipoles in the presence of external field are investigated. In
section \ref{3} we study the noncommutative Aharonov-Casher effect,
in section \ref{4} we extend the study for the He-McKellar-Wilkens
phase in a noncommutative space-time. In section \ref{6} we discuss
this quantum phase considering the momentum-momentum
non-commutativity or phase space non-commutativity. Finally, in
section \ref{5} the conclusions are presented.

\section{The Nonrelativistic Limit}\label{1}
Now, we consider the relativistic quantum dynamics of a single
neutral spin half particle with non-zero magnetic and electric
dipole moving in external electromagnetic field, which is described
by the following equation (we used $\hbar=c=1$).

\be\label{eqc1}
[i\gamma_\mu\partial^\mu+\frac12\mu\sigma_{\alpha\beta}F^{\alpha\beta}-
\frac{i}{2}d\sigma_{\alpha\beta}\gamma_5F^{\alpha\beta}-m]\psi=0\;,
\ee where $\mu$ is the magnetic dipole moment and $d$ is the
electric dipole moment. We use the following convention for field
strength\cite{gross}:

\begin{eqnarray*}
F^{\mu\nu}=\{\vec{E},\vec{B}\}\;,\qquad F^{\mu\nu}=-F^{\nu\mu}\;,\\
F^{i0}=E^i\;,\qquad F^{ij}=-\epsilon_{ijk}B^k\;,
\end{eqnarray*}
\begin{equation}
F^{\mu\nu}=\left[\begin{array}{cccc}
0 & -E^x & -E^y & -E^z\\
E^x & 0 & -B^z & B^y\\
E^y & B^z & 0 & -B^x\\
E^z & -B^y & B^x & 0
\end{array}\right]\;,
\end{equation}
\ben
&&\sigma_{0i}=\frac{i}{2}[\gamma_{0}\gamma_{i}-\gamma_{i}\gamma_{0}]=i\gamma_{0}\gamma_{i}=-i\alpha_{i}\;,\nonumber\\&&
\sigma_{ij}=\frac{i}{2}[\gamma_{i}\gamma_{j}-\gamma_{j}\gamma_{i}]=i\gamma_{i}\gamma_{j}=\epsilon_{ijl}\Sigma_{l}\;,
\een
and our $\gamma_5$ choice for convenience is
\be
\gamma_{5}=\left(\begin{array}{cc}
0 & -1 \\
-1 & 0\\ \end{array}\right)\;,
\ee
where $0$ and $-1$ are
corresponding $2\times2$ matrices \cite{silenko}. Hence, we may
write the equation (\ref{eqc1}) as \be\label{EM}
[i\gamma^\mu\partial_\mu+\mu(i\vec{\alpha}\cdot\vec{E}-\vec{\Sigma}\cdot\vec{B})
-id(i\vec{\alpha}\cdot\vec{E}-\vec{\Sigma}\cdot\vec{B})\gamma_5-m]\psi=0\;,
\ee
with the Dirac matrices given by:
\begin{eqnarray*}
&\hat{\beta}=\gamma^{0}=\left(\begin{array}{cc}
1 & 0 \\
0 & -1\\ \end{array}\right)\;,\quad\gamma^{j}=\left(\begin{array}{cc}
0 & \sigma^{j} \\
-\sigma^{j} & 0\\ \end{array}\right)\;,\quad\vec{\alpha}=\hat{\beta}\vec{\gamma}=\left(\begin{array}{cc}
0 & \vec{\sigma} \\
\vec{\sigma} & 0\\ \end{array}\right)\;,&\\
&\vec{\Sigma}=\left(\begin{array}{cc}
\vec{\sigma} & 0 \\
0 & \vec{\sigma}\\ \end{array}\right)\;,\quad\vec{\Pi}=\hat{\beta}\vec{\Sigma}=\left(\begin{array}{cc}
\vec{\sigma} & 0 \\
0 & -\vec{\sigma}\\ \end{array}\right)\;.&
\end{eqnarray*}
where $\sigma^{j}$ are the  Pauli matrices obeying the  relation $\{\sigma^i\sigma^j+\sigma^j\sigma^i\}=-2g^{ij}$.
Then, the Hamiltonian given by equation(\ref{EM}) is reduced to the form
\be
i\frac{\partial}{\partial t}\psi=H\psi=[\vec{\pi}\cdot\vec{\alpha}+\mu\,\vec{\Pi}\cdot\vec{B}+d\,\vec{\Pi}\cdot\vec{E}
+\hat{\beta}m]\psi;,
\ee
where $\vec{\pi}=-i(\vec{\nabla}+\mu\,\hat{\beta}\vec{E}-d\,\hat{\beta}\vec{B})$.

From now on, we will use the Foldy-Wouthuysen method. This a very convenient method of description of 
the relativistic particle interaction with an external field and of transition to the semiclassical 
description is the Foldy-Wouthuysen transformation\cite{fw}. The Foldy-Wouthuysen representation provides 
the best opportunity for the transition to the classical limit of relativistic quantum mechanics.
The Hamiltonian in equation (\ref{EM}) takes the form
\be
\hat{H} = \hat{\beta}( m + \hat{\epsilon}) + \hat{O}\;,
\ee
where $\hat{\epsilon}=\mu\,\vec{\Pi}\cdot\vec{B}+d\,\vec{\Pi}\cdot\vec{E}$ and $\hat{O} = \vec{\pi}\cdot\vec{\alpha}$ are the even and odd terms in the Hamiltonian. Hence we introduce the transformation
\be
\hat{H}^{'}= e^{i\hat{S}}(\hat{H} - i\partial_{0})e^{-i\hat{S}}\;,
\ee
where $\hat{S}$ is a hermitian matrix. The purpose is to minimize the odd part of the Hamiltonian, or even to make it vanishing. Thus, we have 
\ben
\hat{H}^{'}&=& \hat{H}+i[\hat{S},\hat{H}]-\frac{1}{2}[\hat{S},[\hat{S},\hat{H}]]-
\frac{1}{6}i[\hat{S},[\hat{S},[\hat{S},\hat{H}]]]\nonumber \\&&+\frac{1}{24}[\hat{S},[\hat{S},[\hat{S},\hat{\beta}m]]]+...\;,
\een
where $\hat{\epsilon}$ and $\hat{O}$ above obey the relations $\hat{\epsilon}\hat{\beta}=\hat{\beta}\hat{\epsilon}$ and $\hat{O}\hat{\beta}=-\hat{\beta}\hat{O}$.

For nonrelativistic particles in an electromagnetic field, the FW transformation can be
performed with the operator$\hat{S}=-\frac{i}{2m}\hat{\beta}\hat{O}$, such that we have
\be
\hat{H}^{'} = \hat{\beta} m + \hat{\epsilon}^{\;'} + \hat{O}^{'} \;,
\ee
where $\hat{O}^{'}$ is at the order of $\frac{1}{2m}$ , and we calculate the second-order FW transformation with $\hat{S}^{\;'}=-\frac{i}{2m}\hat{\beta}\hat{O}^{'}$. This yields
\be
\hat{H}^{''} = \hat{\beta} m + \hat{\epsilon}^{\;'} + \hat{O}^{''}\;,
\ee
where $\hat{O}^{''}\approx\frac{1}{m^{2}}$ .  After that, the third FW approximation with  $\hat{S}^{\;''}=-\frac{i}{2m}\hat{\beta}\hat{O}^{''}$ makes the odd part of the nonrelativistic expansion vanishing; so  finally we find the usual result
\ben
\hat{H}^{'''}&\cong& \hat{\beta}m + \hat{\epsilon}^{\;'} \nonumber\\&
=&\hat{\beta}(m + \frac{1}{2m}\hat{O}^{2}-\frac{1}{8 m^{3}}\hat{O}^{4})+\hat{\epsilon}
- \frac{1}{8 m^{2}}[\hat{O},[\hat{O},\hat{\epsilon}]]\;.\label{h3rd}
\een

After replacing $\hat{\epsilon}$ e $\hat{O}$ in (\ref{h3rd}), we will consider only the terms up to order 1/m. 
We obtain the following Hamiltonian
\ben
\hat{H}^{'''}\approx\hat\beta\Bigg[m&-&\frac{1}{2m}\left(\vec{\nabla}-i\mu\hat\beta(\vec{\Sigma}\times\vec{E})+id\hat\beta(\vec{\Sigma}\times\vec{B})\right)^2 -\frac{\mu^2E^2}{2m}-\frac{d^2\vec{B}^2}{2m}\nonumber\\
&-&\frac{\mu\hat\beta}{2m}\vec\nabla\cdot\vec{E}+\frac{d\hat\beta}{2m}\vec\nabla\cdot\vec{B}\Bigg]+\mu\vec{\Pi}\cdot\vec{B}+d\vec{\Pi}\cdot\vec{E}
\;.\label{eq10}
\een
The equation (\ref{eq10}) is the nonrelativistic    quantum Hamiltonian for four-components fermions. However, for several applications in low energies in nonrelativistic quantum mechanics the two-components spinor field is considered, we may write (\ref{eq10}) for two-components fermions in the form
\ben
\hat{H}^{'''}&\approx& m-\frac{1}{2m}\left(\vec{\nabla}-i(\vec{\mu}\times\vec{E})+i(\vec{d}\times\vec{B})\right)^2 -\frac{\mu^2E^2}{2m}-\frac{d^2\vec{B}^2}{2m}\nonumber\\
&&-\frac{\mu}{2m}\vec\nabla\cdot\vec{E}+\frac{d}{2m}\vec\nabla\cdot\vec{B}+\vec\mu\cdot\vec{B}+\vec{d}\cdot\vec{E}
\;,\label{eqq10}
\een
where $\vec\mu=\mu\vec\sigma$ and $\vec d=d\vec\sigma$, and $\vec{\sigma}=(\sigma_1,\sigma_2,\sigma_3)$; where $\sigma_i$ $(i=1,2,3)$ are the $2\times2$ Pauli matrices.
The Hamiltonian (\ref{eqq10}) describes a system formed by a neutral particle, that possesses permanent electric and magnetic dipole moments, in the presence of electric and magnetic fields. Several topological and geometrical effects may be investigated by changing the fields-dipole configuration \cite{mac,wil,wei}.
\section{Nonrelativistic Quantum Dynamics of Dipoles}\label{2}
Considering the nonrelativistic quantum dynamics of a particle corresponding to the  Hamiltonian (\ref{eq10}) which describes several  physical situations such as: the Aharonov--Casher effect $\mu\neq 0$ and $d=0$, the He-McKellar-Wilkens effects $\mu = 0$ and $d\neq0$,  and the Anandan phase in general case $\mu\neq 0$ and $d\neq 0$. It is obvious that all this effects occur in  specific field-dipole configuration.
We analyze the quantum dynamics of a particle governed by the Hamiltonian (\ref{eqq10}). We consider that the electric and magnetic fields, whose particle is immersed, are cylindrically radial \cite{mac,wil,wei,prlan,plaan,furtpra}. The electric and magnetic dipoles are aligned in $z$-direction. The Hamiltonian (\ref{eqq10}) that describe the electric and magnetic dipoles in an external electric and magnetic fields can be written in following way:
\ben \label{eqq10b}
H= -\frac{1}{2m}\left(\vec{\nabla}-b_{\mu} \right)^2  +b_{0}\;,\label{eqq10a}
\een
here the interaction with the electric and magnetic fields is similar as if it is minimally coupled to a non-Abelian gauge field with potential $b_{\mu}$, where
\begin{equation}
b_{0}= -\frac{\mu^2E^2}{2m}-\frac{d^2\vec{B}^2}{2m}-\frac{\mu}{2m}\vec\nabla\cdot\vec{E}+\frac{d}{2m}\vec\nabla\cdot\vec{B}+\vec\mu\cdot\vec{B}+\vec{d}\cdot\vec{E}\;,
\end{equation}
and $b_{i}=(\vec{\mu}\times\vec{E})+(\vec{d}\times\vec{B})$.
The first two terms in $b_{0}$ can be considered as an external potential and not contribute in the study of the geometric phase \cite{wei}. Observe that the potential $b_{0}$, which depends on $E^{2}$ and $B^{2}$, represents a local influence to the wave function. We are
interested in to study the asymptotic states for the  dynamics. Then we will not consider this term, because it represent a local effect \cite{wei}. Also  Anadan \cite{plaan} has demonstrated that terms of the order $O(E^{2})$ and $O(B^{2})$ can be neglected in the study of geometric phase. 
The last four terms in  potential  $b_{0}$
in the Hamiltonian (\ref{eqq10b})  give null contribution in the
dynamic of this dipole in the fields configuration since then the dipoles are aligned with
$z$-direction \cite{plaan,prlan,furtpra}.  Assuming that the particle moving in plane $x-y$ , in presence of the
external electric and magnetic fields \cite{wei}. We also suppose that the fields
generated by the source are radially distributed in the space.  Now, we consider the commutative space version of this geometric phases. So that the only terms that contribute to geometric phase in (\ref{eqq10b}) are 
\ben
\hat{H}=-\frac{1}{2m}\left(\nabla-i(\vec{\mu}\times\vec{E})+i(\vec{d}\times\vec{B})\right)^2 \label{hamil} -\frac{\mu^2E^2}{2m}-\frac{d^2\vec{B}^2}{2m}\;,
\een
the other terms of (\ref{eq10}) do not contribute to quantum phase due to choice of specific dipole-field configurations. The terms, in dynamical part of Hamiltonian, yields no force on the particle, while in quantum mechanics it affects the wave function of the particles by attaching  to it a nondispersive geometric phase. The momentum operator can be written as
\ben
k_{i}=mv_{i}=(p_{i}-(\vec{\mu}\times\vec{E})_{i}+(\vec{d}\times\vec{B})_{i})\;.
\een
The Schr\"odinger equation for this problem takes the form
\ben\label{schn}
\left(-\frac{1}{2m}(\nabla-i(\vec{\mu}\times\vec{E})+i(\vec{d}\times\vec{B}))^2  -\frac{\mu^2E^2}{2m}-\frac{d^2\vec{B}^2}{2m}\right)\Psi=E\Psi\;,
\een
To obtain the quantum phase we use the following ansatz
\ben
\Psi=\Psi_{0}e^{\phi}\;,
\een
where $\Psi_{0}$ is solution of following equation
\ben
\left(-\frac{1}{2m}\nabla^2 -\frac{\mu^2E^2}{2m}-\frac{d^2\vec{B}^2}{2m}\right) \Psi_{0}=E\Psi_{0}\;,
\een
and the phase $\phi$ is given by
\ben\label{anan}
\phi=i\oint[(\vec{\mu}\times\vec{E})-(\vec{d}\times\vec{B})]\cdot dr\;,
\een
this phase was studied by Anandan \cite{prlan}. It is a nondispersive effect due the independence on particle velocity \cite{bud}. Considering $d=0$ in (\ref{anan}), we have the Aharonov-Casher geometric phase. On the other hand, in the case $d\neq0$ and $\mu=0$ in (\ref{anan}) we have the He-McKellar-Wilkens phase,
\ben\label{hmwc}
\phi_{HMW}=i\oint[-(\vec{d}\times\vec{B})]\cdot dr\;,
\een
that is  usually known as topological phase but really it  is a geometric phase \cite{pes2}.
\section{Noncommutative Quantum Dynamics of Dipoles}\label{21}
The usual noncommutative space canonical variables satisfy  the following commutations relations
\begin{eqnarray}\label{relat}
[\hat{x_{i}},\hat{x_{j}}]=i\Theta_{ij}\;, \quad
[\hat{p_{i}},\hat{p_{j}}]=0\;,\quad
[\hat{x_{i}},\hat{p_{j}}]=i\delta_{ij}\;,
\end{eqnarray}
where $\hat{x_{i}}$ and $\hat{p_{i}}$ are momentum and coordinate operators in a noncommutative space.
The time-independent Schr\"odinger equation in the noncommutative (NC) space can be written in the form
\be
H(x,p)\star\psi=E \psi\;,\label{ECNC}
\ee
where $H(x,p)$ is usual Hamiltonian and the Moyal-Weyl product (or star-product) is given by
\be\label{starprod}
(f \star g)(x)=\exp[\frac{i}{2}\Theta_{ij}\partial_{x_i}\partial_{x_j}]f(x_i)g(x_j)\;,
\ee
here $f(x)$ and $g(x)$ are arbitrary functions. On the NC space the Moyal-Weyl product may be replaced by a Bopp's shift \cite{Bopp's}, i.e. the Moyal-Weyl product can be changed in the ordinary product by replacing $\text{H}(x,p)$ with $\text{H}(\hat{x},\hat{p})$. This  approach has been used by Li et al \cite{5}. Hence, the Schr\"odinger equation can be written in the form
\be
H(\hat{x}_i,\hat{p}_i)=H(x_{i}- \frac{1}{2}\Theta_{ij}p_{i},p_{i})\psi=E \psi\;,\label{MQNC}
\ee
where $x_{i}$ and $p_{i}$ are the generalized position and momentum coordinates in the usual quantum mechanics. Therefore, the equation (\ref{MQNC}) is then actually defined on the commutative space, and the NC effect may be calculated by the terms that contain $\Theta$. Note that $\Theta$ in quantum mechanics may be taken as a perturbation considering $\Theta_{ij}<< 1$.

When we have the presence of electric and magnetic fields as in (\ref{schn}), the equation (\ref{ECNC}) becomes
\ben\label{eqc09}
\left(-\frac{1}{2m}\left(\vec\nabla-i(\vec{\mu}\times\vec{E})+i(\vec{d}\times\vec{B})\right)^2 -\frac{\mu^2E^2}{2m}-\frac{d^2\vec{B}^2}{2m} \right)\star\Psi=E\Psi\;,
\een
To map the equation (\ref{eqc09}) from NC space to commutative space, we replace $x_i$ and $p_i$ by a Bopp's shift \cite{Bopp's}, as well as the fields $E_i$ and $B_i$ that will be replaced with a shift in the form
\be\label{eqcc25}
(\vec{\mu}\times \vec{E})\rightarrow(\vec{\mu}\times \vec{E})+\frac{i}{2}\Theta_{lm}(\vec{\kappa}-(\vec{\mu}\times \vec{E}))_{l}\partial_{m}(\vec{\mu}\times \vec{E})\;,
\ee
and 
\be\label{eqcc26}
(\vec{d}\times \vec{B})\rightarrow (\vec{d}\times \vec{B})+\frac{i}{2}\Theta_{lm}(\vec{\kappa}-(\vec{d}\times \vec{B}))_l\partial_{m}(\vec{d}\times \vec{B})\;,
\ee
where the $\kappa_{l}$ is the eigenvalue of momentum operator in the presence of the electric or magnetic field on NC space, and defined as 
\be\label{eqcc27}
(p_i-(\vec{\mu}\times\vec{E})+(\vec{d}\times\vec{B}))\star\psi=\kappa_i\psi\;,
\ee
where, $\kappa_i=mv_i$ and $v_i$ is the ordinary gradient. The relations (\ref{eqcc25}) and (\ref{eqcc26}) may be obtained in the same form by Taylor expansion up to first order of (\ref{starprod}), for example, let us take the magnetic dipole case:
\ben
\Big((\vec{\mu}\times\vec{E})\star\psi\Big)(x)&=&\exp{\left[\frac{i}{2}\Theta_{ij}\partial_{x_i}\partial_{x_j}\right]}(\vec{\mu}\times\vec{E})(x_i)\psi({x_j})\nonumber\\
&=&(\vec{\mu}\times\vec{E})\psi+\frac{i}{2}\Theta_{ij}\partial_{i}(\vec{\mu}\times\vec{E})\partial_{j}\psi
\een
from (\ref{eqcc27}) we have
\be\label{eqcc28}
\partial_i\psi=(\kappa-(\vec{\mu}\times\vec{E}))_i\psi\;.
\ee
Therefore, using (\ref{eqcc28}) into (\ref{eqcc27}) we obtain (\ref{eqcc25}). In the same way we may obtain (\ref{eqcc26}). Thus, the NC equation (\ref{eqc09}) mapped on commutative space is
\begin{eqnarray}
-\frac{1}{2m}\Big(\vec{\nabla}-i (\vec{\mu}\times \vec{E})-\frac{i}{2}\Theta_{lm}(\kappa_{l}-(\vec{\mu}\times \vec{E})_l)\partial_{m}(\vec{\mu}\times \vec{E})+\nonumber\\+ i(\vec{d}\times \vec{B})
+\frac{i}{2}\Theta_{lm}(\kappa_{l}-(\vec{d}\times \vec{B})_l)\partial_{m}(\vec{d}\times \vec{B}) \Big)^{2}\psi=E\psi\;.\label{completescrodinger}
\end{eqnarray}
In the same way of the usual quantum mechanics, the solution for (\ref{completescrodinger}) may be written as
\be
\psi=\psi_0\exp(\phi)\;,
\ee
where $\psi_{0}$ is a solution of the Schr\"odinger equation in the absence of electric and magnetic  fields and $\phi$ is the Anandan's geometric phase given in the form
\ben
\phi&=&i\oint[(\vec{\mu}\times \vec{E})-(\vec{d}\times \vec{B})]\cdot d\vec{r}+\nonumber\\ &&+\frac{i}{2}\Theta_{lm}\oint\{(\kappa-(\vec{\mu}\times \vec{E}))_l\partial_{m}(\vec{\mu}\times \vec{E})
-(\kappa-(\vec{d}\times \vec{B}))_l\partial_{m}(\vec{d}\times \vec{B})\}\cdot d\vec{r}\;.\label{anc}
\een
The first term of the integral in the equation (\ref{anc}) is the usual Anandan's phase in the commutative quantum mechanics. The other terms are the corrections due to NC effects. In the three-dimensional commutative space, we define the vector $\theta=(\theta_1,\theta_2,\theta_3)$ with $\Theta_{ij}=\epsilon_{ijk}\theta_k$. Thus we rewrite the total phase (\ref{anc}) in the form
\ben
\phi&=&i\oint[(\vec{\mu}\times \vec{E})-(\vec{d}\times \vec{B})]\cdot d\vec{r}+\frac{i}{2}m\oint\vec{\theta}\cdot[\vec{v}\times\vec{\nabla}(\vec{\mu}\times \vec{E})_i] d{r}_i-\nonumber\\
&&-\frac{i}{2}m\oint\vec{\theta}\cdot[(\vec{\mu}\times \vec{E})\times\vec{\nabla}(\vec{\mu}\times  \vec{E})_i]d{r}_i-\frac{i}{2}m\oint\vec{\theta}\cdot[\vec{v}\times\vec{\nabla}(\vec{d}\times \vec{B})_i]d{r}_i+ \nonumber\\
&&+\frac{i}{2}m\oint\vec{\theta}\cdot[(\vec{d}\times \vec{B})\times\vec{\nabla}(\vec{d}\times \vec{B})_i] d{r}_i\;.
\label{phasedip}\een
The phase (\ref{phasedip}) is noncommutative version of a nonrelativistic quantum Anandan's phase. Notice that the dependence of  phase in the electric and magnetic field. Other properties of Eq. (\ref{phasedip}) is that geometric phase depend of the velocity of the particle. The noncomutativity of space introduces this dependence in the phase. In the next section we wil discuss some special limits of this geometric phase.
\section{Noncommutative Aharonov--Casher Effect }\label{3}
First we consider the case in the expression (\ref{anc}) where $d=0$. In this case, we obtain the  Aharonov--Casher (AC) phase  given by
\be
\phi_{AC}=i\oint[\vec{\mu}\times \vec{E}+\frac{1}{2}\Theta_{lm}(\kappa_{l}-(\vec\mu\times \vec{E})_l)\partial_{m}(\vec{\mu}\times \vec{E})]\cdot d\vec{r}\;.\label{acph}
\ee
The first term in the integral in the equation (\ref{acph}) is the usual AC phase in the commutative quantum mechanics. The second term is the NC correction to the AC phase. In the three-dimensional commutative space, we define the vector $\theta=(\theta_1,\theta_2,\theta_3)$ with $\Theta_{ij}=\epsilon_{ijk}\theta_k$. Thus we rewrite the total phase (\ref{acph}) in the form
\be
f=i\oint(\vec{\mu}\times \vec{E})\cdot d\vec{r}+\frac{i}{2}m\oint\vec{\theta}\cdot[\vec{v}\times\vec{\nabla}\cdot(\vec{\mu}\times \vec{E})]\cdot d\vec{r}
-\frac{i}{2}m\oint\vec{\theta}\cdot[(\vec{\mu}\times \vec{E})\times\vec{\nabla}\cdot(\vec{\mu}\times \vec{E})]\cdot d\vec{r}\;.
\ee
This geometric phase is the same obtained in \cite{mirza,5} in the relativistic case. Note that this is a dispersive geometric phase that depends on the velocity of the particle \cite{bud}.
\section{ The noncommutative He--McKellar--Wil\-kens effects}\label{4}
Now let us analyze the particular case of (\ref{anan}) in the noncommutative situation. This case is the noncommutative He-McKellar-Wilkens quantum phase in nonrelativistic limit. The He and McKellar and Wilkens independently \cite{mac, wil} have demonstrated that a quantum dynamics of a electric dipole in the presence of a radial magnetic field exhibits a geometric phase. The way to obtain the NC He--McKellar--Wilkens (HMW) effect is similar to the AC case. Taking $\mu=0$ in the Pauli's term in the equation (\ref{eq10}). Hence, we find the following NC Schr\"odinger equation for the electric dipole in the presence a magnetic field. Applying this limit in the phase (\ref{anc}) we obtain the following expression
\be
\phi_{HMW}=-i\oint[\vec{d}\times \vec{B}+\frac{1}{2}\Theta_{lm}(\kappa_{l}-(\vec{d}\times \vec{B})_l)\partial_{m}(\vec{d}\times \vec{B})]\cdot d\vec{r}\;.\label{hmwph}
\ee

The first term in (\ref{hmwph}) is the commutative usual HMW quantum phase. The second term is the NC correction to HMW phase. In the same way as in the AC case, in the three-dimensional commutative space we define the vector $\theta=(\theta_1,\theta_2,\theta_3)$ with $\Theta_{ij}=\epsilon_{ijk}\theta_k$. Thus we rewrite the total phase (\ref{hmwph}) in the form
\ben
\phi_{HMW}&=&-i\oint(\vec{d}\times \vec{B})\cdot d\vec{r}-\frac{i}{2}m \oint\vec{\theta}\cdot[\vec{v}\times\vec{\nabla}(\vec{d}\times \vec{B})]\cdot d\vec{r}+\nonumber\\
&&+\frac{i}{2}m\oint\vec{\theta}\cdot[(\vec{d}\times \vec{B})\times\vec{\nabla}\cdot(\vec{d}\times \vec{B})]\cdot d\vec{r}\;.
\een
This equation gives the expression of the noncommutative version of He-MacKellar-Wilkens effect.

\section{Noncommutative Dynamics of Dipoles in Phase Space}\label{6}
In previous section we discuss the noncommutative version of geometric phase in the quantum dynamics of a neutral particle that possesses permanent electric and magnetic dipole moments. Now we will discuss the case where we are taking into account momentum-momentum noncommutativity. The Bose-Einstein statistics in noncommutative quantum mechanics requires both space-space and momentum-momentum noncommutativity \cite{nair,zhang,5,kangepjc,li3}. This formulation has been denominated of phase space noncommutativity. In this case, the momentum commutation relation in (\ref{relat}) is replaced by
\be
[\hat{p_{i}},\hat{p_{j}}]=i\bar{\Theta}_{ij}\;,
\ee
where $\bar{\Theta}$ is the antisymmetric matrix, its elements represent the non-commutativity of the momenta.
Thus the Schr\"odinger equation (\ref{ECNC}) is written in the form
\be\label{ncann1}
-\frac{1}{2m}\left(\nabla-i(\vec{\mu}\times\vec{E})+i(\vec{d}\times\vec{B})\right)^2 \star\Psi=E\Psi\;.
\ee
On noncomutative phase space the star product can be replaced by a generalized Bopp's shift \cite{Bopp's}, in this way the star product can be changed into ordinary product by shifting coordinates $x_{\mu}$  and momenta $p_{\mu}$ by
\be
\hat{x_{i}}=\lambda x_{i} -\frac{1}{2\lambda}\Theta_{ij}p_{j}\;,
\ee
and
\be
\hat{p_{i}}=\lambda p_{i} -\frac{1}{2\lambda}\bar{\Theta}_{ij}x_{j}\;,
\ee
where the scale factor $\lambda$ is an arbitrary constant parameter.
The fields in equation  change according to the formula (\ref{MQNC}) and assume the following form
\be
(\vec{\mu}\times \vec{E})\rightarrow \lambda (\vec{\mu}\times \vec{E})+\frac{i}{2\lambda}\Theta_{lm}(\kappa_{l}-(\vec{\mu}\times \vec{E})_l)\partial_{m}(\vec{\mu}\times \vec{E})\;,
\ee
and the magnetic field term changes for the form
\be
(\vec{d}\times \vec{B})\rightarrow \lambda(\vec{d}\times \vec{B})+\frac{i}{2\lambda}\Theta_{lm}(\kappa_{l}-(\vec{d}\times \vec{B})_l)\partial_{m}(\vec{d}\times \vec{B})\;.
\ee
Now the Schr\"odinger equation for the neutral particle becomes
\ben
-\frac{1}{2m}\Big(\lambda \vec{\nabla}+ \frac{i}{2\lambda}\bar{\Theta}_{ij}x_{i} -i \lambda(\vec{\mu}\times \vec{E})+\frac{1}{2\lambda}\Theta_{lm}(\kappa_{l}-(\vec{\mu}\times \vec{E})_l)\partial_{m}(\vec{\mu}\times \vec{E})\nonumber\\ +i \lambda (\vec{d}\times \vec{B})-\frac{1}{2\lambda}\Theta_{lm}(\kappa_{l}-(\vec{d}\times \vec{B})_l)\partial_{m}(\vec{d}\times \vec{B}) \Big)^{2}\psi=E\psi\;.\label{completescrodingerps}
\een
We can rewrite the Schr\"odinger equation in the following form
\ben\label{com}
-\frac{1}{2m'}\Big(\vec{\nabla}+ \frac{i}{2\lambda^{2}}\bar{\Theta}_{ij}x_{i} -i(\vec{\mu}\times \vec{E})+\frac{1}{2\lambda^{2}}\Theta_{lm}(\kappa_{l}-(\vec{\mu}\times \vec{E})_l)\partial_{m}(\vec{\mu}\times \vec{E})\nonumber\\ +i(\vec{d}\times \vec{B})-\frac{1}{2\lambda^{2}}\Theta_{lm}(\kappa_{l}-(\vec{d}\times \vec{B})_l)\partial_{m}(\vec{d}\times \vec{B}) \Big)^{2}\psi=E\psi\;,\label{completescrodingerps1}
\een
where $m'=m/\lambda$. In the same way of the usual quantum mechanics, the solution for (\ref{completescrodingerps1}) may be written as
\be
\psi=\psi_0\exp(\phi_{PS})\;,
\ee
where $\psi_{0}$ is a solution of the Schr\"odinger equation for a particle of mass $m'$ in the absence of electromagnetic field and $\phi_{PS}$ is the Anandan's geometric phase in non-commutative phase space given in the form
\ben
\phi_{PS}&=&i\oint\{\vec{\mu}\times \vec{E}-\frac{1}{2\lambda^{2}}\Theta_{lm}(\kappa_{l}-(\vec{\mu}\times \vec{E})_l)\partial_{m}(\vec{\mu}\times \vec{E})\}\cdot d\vec{r}-\nonumber\\
&&-i\oint\{\vec{d}\times \vec{B}-\frac{1}{2\lambda^{2}}\Theta_{lm}(\kappa_{l}-(\vec{d}\times \vec{B})_l)\partial_{m}(\vec{d}\times \vec{B})\}\cdot d\vec{r}-\nonumber\\
&&-\frac{i}{2\lambda^{2}}\oint \bar{\Theta}_{ij}x_{j}dx_{i}\;.\label{phasegerps}
\een

The previous expression (\ref{phasegerps}) has a contribution arisen due to a quantum phase in commutative space, other contribution due to a noncommutative space and one more contribution due to a noncommutative phase space. We can write the quantum phase (\ref{phasegerps}) in the following form
\ben\label{phasepscon}
\phi_{PS}=\phi_{AP} + \phi_{NCS} +  \phi_{NCPS}\;,
\een
where the $\phi_{AP}$ and the $\phi_{NCS}$ are the Anandan's phase contribution and the contribution due to the space-space noncommutativity to the general dipole phase in the expression (\ref{phasegerps}). The term $\phi_{NCPS}$ is the contribution due to noncommutativity of the momenta is given by
\ben
\phi_{NCPS}&=&-\frac{i}{2\lambda^{2}}\oint \bar{\Theta}_{ij}x_{j}dx_{i}+
\frac{1-\lambda^{2}}{2\lambda^{2}}\oint[\Theta_{lm}(\kappa_{l}-(\vec{\mu}\times \vec{E})_l)\partial_{m}(\vec{\mu}\times \vec{E})]\cdot d\vec{r}\nonumber\\&&-\frac{1-\lambda^{2}}{2\lambda^{2}}\oint [\Theta_{lm}(\kappa_{l}-(\vec{d}\times \vec{B})_l)\partial_{m}(\vec{d}\times \vec{B})]\cdot d\vec{r}\;.\label{pnps}
\een
This is the contribution to noncommutative geometric phase due to a noncommutativity in phase space. In this way we can write the He--McKellar--Wilkens phase in noncommutative phase space with a particular case where $\vec\mu=0$, and this expression is given by
\be\label{geomncps}
\phi_{HMWPS}=-\frac{i}{2\lambda^{2}}\oint \bar{\Theta}_{ij}x_{j}dx_{i}-\frac{1-\lambda^{2}}{2\lambda^{2}}\oint [\Theta_{lm}(\kappa_{l}-(\vec{d}\times \vec{B})_l)\partial_{m}(\vec{d}\times \vec{B})]\cdot d\vec{r}\;.
\ee
Therefore, we obtain the contribution due to NC phase space to the He--McKellar--Wilkens phase. We can see that this phase depends on the magnetic field and also on the velocity of the particle. The first term in (\ref{geomncps}) is similar in appearance to the spin factor that occur in the partition function of spinning particle. The connection of this spin factor with the geometric phase was investigated by Kalhede et al \cite{kalhede} and Levay \cite{lev}. This similarity with spin factor and the physical implications of this terms is topic of a future contribution.
\section{Concludings  remarks}\label{5}
In this paper, we study the nonrelativistic quantum dynamics of a neutral particle, that possesses permanent electric and magnetic dipole moments, in the presence of electric and magnetic external fields. We use the Foldy--Wouthuysen expansion to make transition to the classical limit of relativistic to nonrelativistic quantum mechanics . In this  limit, we investigate the Aharonov-Casher and the He-McKellar-Wilkens effects in the noncommutative coordinates space. Here, we replace the $\star$-product by the Bopp's shift \cite{Bopp's} in the field terms, and then we obtain to AC and HMW quantum phases with the NC corrections. We obtain the noncommutative Anandan's phase and demonstrate that this is a geometric dispersive phase.  Usually, a geometric phase is a local effect, while a topological
phase is nonlocal. Peshkin and Lipkin \cite{pes2} have shown that the Aharonov-Bohm effect is nonlocal, because its value depends upon a physical quantity in a region outside the closed path. It is a topological effect. The Aharonov-Bohm phase is proportional to the winding number of the path around the flux. It is a topological invariant and this phase depends on topology not on distance; hence it must be nonlocal. Therefore there are no electromagnetic fields along the paths of the charged particle and there are no change of physical quantities. They
remarked that in the case of the Aharonov-Casher  effect there are fields along the paths of the beams; then they concluded that
the Aharonov-Casher effect is local due to the local interactions and is nontopological one because the phase shift depends on the local fields along the paths. In contrast with the topological phase, the geometric phase, in general, is a local effect because it depends on the geometry and topology in the space of parameters but not on the topology in spacetime. Here the quantum phase depends of the fields and the velocity of the particle. This fact characterize the noncomutative Anandan's quantum phase with a geometric phase due to dependence in the fields and dispersive because of the dependence on the velocity \cite{bud}. The NC Aharonov-Casher is obtained with a limit case of (\ref{phasedip}) and agree with the results in the literature \cite{mirza,kangepjc}. The NC version  He-McKellar-Wilkens phase is calculated for the NC quantum dynamics of electric dipoles and is a geometric dispersive phase. The noncommutative phase space version  of Anandan's phase and He-McKellar-Wilkens phase is obtained in this paper and we conclude they are geometric dispersive phases.

\acknowledgments

This work was partially supported by CNPq, CAPES/PROCAD, CNPQ/FINEP/PADCT and PRONEX/CNPQ/FAPESQ .We thank  A. Yu. Petrov for the critical reading of this manuscript. JRN has been partially supported by FAPESP.


\begin{thebibliography}{99}
\bibliographystyle{unsrt}

\bibitem{aha} Y. Aharonov  and D. Bohm, Phys. Rev. {\bf 115}, 485 (1959).
\bibitem{prl:cham}R. G. Chambers, Phys. Rev. Lett. {\bf 5}, 3 (1960).
\bibitem{pes1}M.  Peshkin  and A. Tonomura, {\it The Aharonov-Bohm Effect} (Springer-Verlag, Berlin, 1989).
\bibitem{cas} Y Aharonov and A. Casher, Phys. Rev. Lett. {\bf 53}, 319, (1984).
\bibitem{cim} A. Cimmino et al., Phys. Rev. Lett. {\bf 63}, 380 (1989).
\bibitem{san} K. Sangster et al., Phys. Rev. Lett. {\bf 71}, 3641 (1993).
\bibitem{mac} X. -G. He and B. H. J. McKellar, Phys. Rev. A, {\bf 47}, 3424 (1993).
\bibitem{wil}M. Wilkens, Phys. Rev. Lett. {\bf 72}, 5 (1994).
\bibitem{wei}H. Wei, R. Han and X. Wei, Phys. Rev. Lett. {\bf 75}, 2071 (1995).
\bibitem{prlan} J. Anandan, Phys. Rev. Lett. {\bf 85}, 1354 (2000).
\bibitem{plaan}J. Anandan, Phys. Lett. A {\bf 138} .
\bibitem{furtpra} C. Furtado and C. A. de Lima Ribeiro, Phys. Rev A {\bf 69}, 064104 (2004).
\bibitem{nekra} M. R. Douglas, N. A. Nekrasov, Rev. Mod. Phys. {\bf 73}, 977 (2001).
\bibitem{con} A. Connes, M. R. Douglas, A Schwarz, JHEP {\bf 9802}, 003 (2001).
\bibitem{sei} N. Seiberg, E. Witten, JHEP {\bf 9909}, 032 (1999).
\bibitem{sus} L. Susskind, "The Quantum Hall fluid and noncommutative Chern-Simons theory" hep-th/0101029.
\bibitem{jela}O. F. Dayi and A. Jellal, J. Math. Phys. {\bf 43}, 4592 (2002).
\bibitem{basu} B. Basu and  Subir Ghosh  Phys.Lett. {\bf A 346} 133 (2005).
\bibitem{1} M. Chaichian, M. M. Sheikh-Jabbari and A. Tureanu, Phys. Rev. Lett. {\bf 86}, 2716 (2001).
\bibitem{2} M. Chaichian, A. Demichev, P. Presnajder, M. M Sheikh-Jbbari and A. Tureanu, Nucl. Phys. {\bf B 611}, 383 (2001).
\bibitem{3} M. Chaichian, P. Presnajder, M. M Sheikh-Jabbari and A. Tureanu, Phys. Lett. {\bf B 527}, 149 (2002)
\bibitem{4} H. Falomir, J. Gamboa, M. Loeve, F. Mendez and J. C. Rojas, Phys. Rev. D {\bf 66}, 045018(2002).
\bibitem{5} K. Li and S. Dulat, Eur. Phys. J. C {\bf 46} 825 (2006).
\bibitem{mirza}B. Mirza and M. Zarei, Eur. Phys J. C {\bf 32} (2004).
\bibitem{kangepjc} K. Li and J. Wang, ``The Topological AC effect on noncommutative phase space" hep-th/0608100.
\bibitem{alavi} S. A. Alavi, Physica Scripta {T7} 366 (2003).
\bibitem{ghos}B. Basu, Subir Ghosh, S. Dhar, Europhys.Lett.76:395,2006.
\bibitem{cal}X Calmet and M. Selvaggi Phys.Rev.D {\bf 74} 037901,(2006).
\bibitem{gross} Franz Gross, {\it Relativistic Quantum Mechanics and Field Theory}, Wiley Classics, 1999.
\bibitem{silenko} A. J. Silenko, Russ. Phys. J. {\bf 48},788, (2005).
\bibitem{fw} L. L. Foldy and S.A. Wouthuysen, Phys. Rev., \textbf{78} 29 (1950).
\bibitem{Bopp's} T. Curtright, D. Fairlie and C. Zachos, Phys. Rev. D{\bf 58} (1998), 025002; L. Mezincescu.
\bibitem{bud}G. Badurek, H. Weinfruter, R. G\"{a}hler, A. Kollmar, S. Wehinger and A. Zeilinger, Phys. Rev. Lett. {\bf 71}, 307 (1993).
\bibitem{li3}K. Li, J. Wang, C. Chen, Mod. Phys Lett. {\bf A 20}, 2165 (2005).
\bibitem{nair}V. P. Nair  and A. P. Polychronakos, Phys. Lett {\bf 505}, 267 (2001).
\bibitem{zhang}J.-Z. Zhang, Phys Lett. {\bf B 584}, 204 (2004).
\bibitem{kalhede} Karlhede . Phys. Rev.  {\bf D 41}  2642 (1990)
\bibitem{lev} Levay J. Math. Phys. {\bf 32} 2347 (1991).
\bibitem{pes2}M.  Peshkin  and H. J. Lipkin, Phys. Rev. Lett. {\bf 74}, 2847 (1995).


\end{thebibliography}
\end{document}